\documentclass[twocolumn,pra,aps,amsmath,amssymb,amsfonts]{revtex4}
\usepackage{graphicx}
\usepackage{bm}
\usepackage{subfigure}
\newcommand{\tst}{\textstyle}

\newcommand{\mrm}{\mathrm}

\begin{document}

\title{Applicability of Modified Effective-Range Theory to positron-atom and positron-molecule scattering}
\author{Zbigniew Idziaszek}
\affiliation{CNR-INFM BEC Center, I-38050 Povo, Italy}
\affiliation{Centrum Fizyki Teoretycznej, Polska Akademia Nauk, 02-668 Warsaw, Poland}
\author{Grzegorz Karwasz}
\affiliation{Instytut Fizyki, Uniwersytet Miko\l aja Kopernika, 87-100 Toru\'n, Poland }

\begin{abstract}
We analyze low-energy scattering of positrons on Ar atoms and N$_2$ molecules 
using Modified Effective-Range Theory (MERT)
developped by O'Malley, Spruch and Rosenberg [Journal of Math. Phys. {\bf 2}, 491 (1961)]. 
We use formulation of MERT based on exact solutions of Schr\"odinger equation
with polarization potential rather than low-energy expansions of phase shifts into momentum series.
We show that MERT describes well experimental data, provided that effective-range expansion is performed both for $s$- and $p$-wave scattering, 
which dominate in the considered regime of positron energies ($0.4$ - $2$ eV). We estimate the 
values of the $s$-wave scattering lenght and the effective range for $e^{+}$-Ar and $e^{+}$-N$_2$ collisions.
\end{abstract}
\maketitle

Electron (and positron) scattering on atoms at very low energies is dominated by polarization forces. Modified effective range theory (MERT) was developed by O'Malley, Rosenberg and Spruch \cite{OMalley,OMalley1} for low-energy scattering of charged particles on neutral polarizable systems in general. O'Malley \cite{OMalley2} applied MERT for electron scattering on noble gases, in particular in the region of Ramsauer- Townsend minimum, using early total \cite{Ramsauer} and momentum transfer \cite{Pack} experimental cross sections. Haddad and O'Malley \cite{Haddad}
used three parameter MERT fit for s-wave phase-shift in electron-argon scattering and Ferch, Granitza and Raith \cite{Ferch} - for Ramsauer minimum in methane. Higher-order terms in MERT, resulting from short-range components of polarizabitity, were introduced by Ali and Fraser \cite{Ali}. MERT analysis for Ne, Ar, Kr up to 1 eV was carefully revisited by Buckman and Mitroy \cite{Buckman}
who used five-parameter fit for $s$-wave and $p$-wave shifts.
 
Applicability of MERT to low-energy positron scattering was already hypothesized in \cite{OMalley1}. However, first measurements of total cross sections for positron scattering at low energies on noble atoms 
come only from seventies 
\cite{Costello,Kaupilla}.

The most systematic data for noble atoms, extending down to 0.3 eV were done in WSU Detroit lab, using positrons from a short-lived C$^{11}$ radionuclid, with about 0.1 eV energy resolution \cite{Kaupilla}. Those data indicated clearly a rise of the cross section in the zero energy limit in gases, like He, Ar, H$_2$, Kr, Xe, CO$_2$, see \cite{Kaupilla}.
Unfortunately, subsequent experiments \cite{Sueoka,Stein} used Ne$^{22}$ source and thick W-vanes positron moderator, thus worsening energy resolution and not allowing to make reliable measurements below 1 eV. To gain in signal, large apertures and strong guiding magnetic fields were used, leading to underestimation of cross sections - some data showed even a fall in the limit of zero energy for highly polarizable targets, like C$_6$H$_6$ \cite{Kimura}.

Only two of the most recent set-ups reached energies below 1 eV with good signal-to-noise ratio. In San Diego annihilation rates in Ar and Xe were measured, showing a steep rise below 1 eV \cite{Marler}. In Trento total cross sections in Ar and N$_2$ were measured \cite{Karwasz1} with angular resolution better by a factor of 30 than in some previous experiments \cite{Sueoka}. Both laboratories confirm the early observations from WSU Detroit on the rise of positron cross sections in the zero-energy limit. Such a rise is also predicted by ab-initio theories 
\cite{McEachran}, see \cite{Karwasz1}
for detailed comparison. A phenomenological attempt to apply MERT-like fit for low-energy cross sections in benzene and cyclohexane was done by Karwasz, Pliszka and Zecca \cite{Karwasz2}.

In the present paper we apply MERT to positron total cross sections on argon and nitrogen, using recent experimental data from Trento \cite{Karwasz1}. 
We use MERT model based on direct solution of Schr\"odinger equation with polarization potential as originally 
proposed by O'Malley, Spruch and Rosenberg \cite{OMalley}.
Differently from earlier works, for the $p$-wave phaseshift we consider not only 
the polarization potential but contribution from a general-type short range interaction. This allows us to extend the MERT applicability for positrons to energies above 1 eV. A clear indication on importance of $p$-wave scattering in this energy range comes from recent differential cross sections measurements in argon \cite{Sullivan}. 
The present model introduces a second MERT parameter for the $p$-wave shift thus developing an approximation with two parameters for both $s$ and $p$-wave phaseshifts. The first parameter is to be interpreted as a scattering length
and the second as an effective range. We compare our MERT model with the expansion into momentum 
series valid at low energies and with ab-initio theories \cite{McEachran,Nakanishi}.

Let us briefly review the effective-range expansion for $1/r^4$ interaction. We divide the interaction potential
between charged particle and neutral atom into the long-range part: 
$V_\mrm{p}(r)= - \alpha e^2/(2 r^4)$ with $\alpha$ denoting the atomic polarizability and $e$ the charge, and the 
short-range part: $V_\mrm{s}(r)$ describing forces acting at distances comparable to the size of atoms.
In the relative coordinate the motion of particles is governed by
\begin{equation}
\label{RadSchr}
\left[\frac{\partial^2}{\partial r^2} + \frac{2}{r} \frac{\partial}{\partial r} - \frac{l(l+1)}{r^2} 
+ \frac{(R^\ast)^2}{r^4} + k^2 \right]\Psi(r) = 0, 
\end{equation}
where $\Psi(r)$ denotes the radial wave function for $l$-th partial wave, $\hbar k$ is the relative momentum of the 
particles, $R^\ast \equiv \sqrt{ \alpha e^2 \mu 
/\hbar^2}$ denotes a typical lenght scale related with the $r^{-4}$ interaction, and $\mu$ denotes the reduced mass.
In Eq.~\eqref{RadSchr} we do not include $V_\mrm{s}(r)$, which is replaced by appropriate boundary condition at $r \rightarrow 0$. The Schr\"odinger equation with polarization potential 
can be solved analytically \cite{Vogt,OMalley,Spector}
(see Appendix for details). At small distances ($r \ll R^{\ast}$) behavior of $\Psi(r)$ is governed by 
\begin{equation}
\label{Psi1}
\Psi(r)  \stackrel{r \rightarrow 0}{\sim} \sin\left(\frac{R^\ast}{r} + \phi\right),
\end{equation}
where $\phi$ is a parameter which is determined by the short-range part of the interaction potential. 
For $r \gg R^{\ast}$, $\Psi(r)$ takes the form of the scattered wave  
\begin{equation}
\label{Psi2}
\Psi(r) \stackrel{r \rightarrow \infty}{\sim} \frac{1}{k r} \sin( {\tst k r - l \frac \pi 2 + \eta_l} ),
\end{equation}
with the phase shift $\eta_l$
\begin{equation}
\label{taneta}
\tan \eta_l = \frac{m^2 - \tan \delta^2 + B \tan \delta (m^2 - 1)}
{\tan \delta (1 - m^2) +  B (1- m^2 \tan^2 \delta)},
\end{equation}
where we use similar notation as in Ref. \cite{OMalley}. Here $B=\tan ({\tst \phi + l \frac \pi 2})$, 
$\delta = \frac \pi 2 (\nu -l -\frac 1 2)$, 
and $m$ and $\nu$ are parameters obtained from the analytic solution
of the Mathieu's differential equation (see Appendix). 
To introduce effective range we expand $B$ around zero energy:
$B(k)= B(0) + \frac12 R_0 R^{\ast} k^2+ \ldots$ \cite{OMalley}.
The second term can be intepreted as correction due to the finite range of the interaction, with $R_0$ representing the effective range. 

In the zero-energy limit expansion of MERT in series of momentum $k$ is useful.
In the particular case of $l=0$, $B(0)$ can be expressed in terms of $s$-wave
scattering length $a_{s}$: $B(0) = - R^{\ast}/a_{s}$, and expansion of $\cot \eta_0$ at $k=0$ yields
\cite{OMalley,Levy}
\begin{align}
q \cot \eta_0(q) = &  - \frac{1}{a} + \frac{\pi}{3 a^2} q + \frac{4}{3 a} \ln \left(\frac q 4 \right) q^2
+ \frac{{R_0}^2}{2 (R^{\ast})^2} q^2 \nonumber \\
& + \left[ \frac \pi 3 + \frac{20}{9a} - \frac{\pi}{3 a^2} - \frac{\pi^2}{9 a^3} 
- \frac{8}{3a} \psi({\tst \frac 32}) \right] q^2 \nonumber \\
& + \ldots
\label{Exp1}
\end{align}
where $a = a_{s}/R^{\ast}$, $q = k R^{\ast}$ and $\psi(\frac 32)$ denotes the digamma function \cite{Abramowitz}.
We apply similar procedure for $p$-wave. In this case, however, 
we expand directly $\tan \eta_1$ given by Eq.~\eqref{taneta}
\begin{align}
\tan \eta_1 = & \frac{\pi q^2}{15} + \frac{q^3}{9 b} - \frac{83 \pi q^4}{23625} - \frac{4}{135 b} 
\ln \left(\frac q 4 \right) q^5 
- \frac{R_1}{18 b^2 R^{\ast}} q^5 \nonumber \\
& + \frac{ 15 \pi - 15 \pi b^2 - 148 b + 120 b \psi(\frac{5}{2})}{2025 b^2} q^5 + \ldots
\label{Exp2}
\end{align}
Here, $b = B(0)$ for $l=1$, and $R_1$ denotes the effective range for $p$-wave.
For higher partial waves we retain only the lowest order term in $k$, which is sufficient to describe the scattering 
in the considered regime of energies
\begin{align}
\tan \eta_l \approx - \frac{\pi q^2}{8(l-\frac12)(l+\frac12)(l+\frac32)}, \quad l \geq 2
\label{Exp3}
\end{align}

Let us turn now to positron scattering. We compare the total cross-section measured in experiment 
for Ar and N$_2$ \cite{Karwasz1} 
with predictions of the theoretical model based on the effective-range expansion. In our approach the 
effects of the short-range potential are included both for $s$- and $p$-wave
giving the leading contribution to the scattering in the considered regime of energies. Thus, our model contains four unknown parameters: the scattering length $a$ and the effective range $R_0$ for $s$-wave, and the zero-energy
contribution $b=B(0)$ and the effective range $R_1$ for $p$-wave. For the investigated regime of positrons energies,
$q=k R^{\ast}$ can take values larger than unity, therefore 
for $s$- and $p$-wave we do not use expansions \eqref{Exp1}-\eqref{Exp2} valid 
for $q \lesssim 1$, but we rather applied the initial formula \eqref{taneta} for the phase shift, performing only 
finite-range expansion for the parameter $B$. In this case, values of $\nu$ and $m$ has to be evaluated numerically, using 
the approach described in the Appendix.

For the calculations we use recent experimental values of the polarizability:
$\alpha=11.23 {a_0}^3$ and  $\alpha=11.54 {a_0}^3$ (atomic units), for Ar and N$_2$, respectively \cite{Olney}.
Table~\ref{Tab} contains values of the characteristic distance $R^{\ast}$ and the characteristic energy 
$E^{\ast} = \hbar^2/(2 \mu {R^{\ast}}^2)$ for the polarization potential, and the values of four parameters: 
$a$, $b$, $R_0$, $R_1$ which were determined by fitting our model to the experimental data.

In the case of N$_2$ the size of the molecule scaled by $R^{\ast}$ is much larger than for Ar, therefore
we restricted our effective-range analysis to lower energies, fitting the model to experimental data with 
$E \lesssim 0.8 E^{\ast}$. In this regime the contribution of the effective-range correction in $p$-wave is rather
small, and one does not get reliable results for this parameter from the fitting procedure. Thus, for N$_2$ we 
considered only three parameters $a$, $b$ and $R_0$ accounting for effects of the short-range part of the
potential.
  
\begin{figure}
	 \includegraphics[width=8cm,clip]{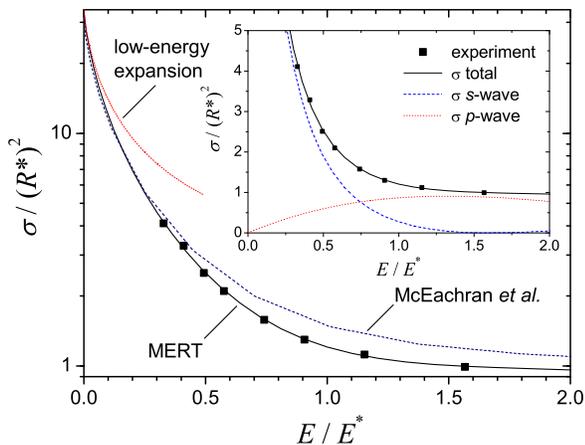}
	 \caption{ Total cross-section for the scattering of positrons on argon versus the energy. Depicted are: experimental
	 data (squares), the theoretical fit based on effective-range expansion (solid line), its low-energy part 
	 given by Eqs.~\eqref{Exp1}-\eqref{Exp3},
	 and the theoretical results of McEachran {\it et al.} (dashed line). The inset shows in addition the $s$- 
	 and $p$-wave cross-sections. Data are scaled by the the characteristic distance $R^{\ast}$ and 
	 the characteristic energy $E^{\ast}$ of the polarization potential.
	 \label{fig:Ar}
	 }
\end{figure}

\begin{figure}
	 \includegraphics[width=8cm,clip]{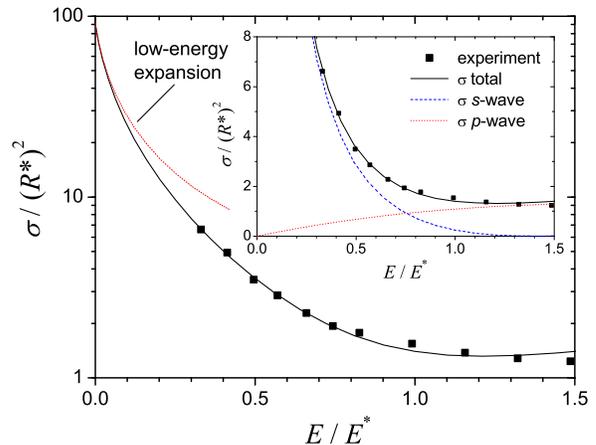}
	 \caption{ The same as Fig.~\ref{fig:Ar} but for the scattering of positrons on N$_{2}$.
	 \label{fig:N2}
	 }
\end{figure}

Fig.~\ref{fig:Ar} shows the experimental data for the total scattering cross-section for Ar as a function of
positron collision energy. They are compared with: the MERT theoretical curve which best fits the experimental data, its
low-energy expansion given by Eqs.~\eqref{Exp1}-\eqref{Exp3}, and  
the results of McEachran {\it et al.} \cite{McEachran}.
The total cross-section is presented in units of $R^{\ast}$, while the energy is scaled by $E^{\ast}$.
In the inset we additionally present contributions of the $s$ and $p$ waves to the total scattering cross-section.
Similar results but for the scattering of positrons on N$_2$ are illustrated in Fig.~\ref{fig:N2}.

We note the good agreement between our model and the results of McEachran {\it et al.} \cite{McEachran} 
at low energies.
Obtained value of the scattering lenght $a_\mrm{s} = -5.58 a_0$ agrees well with 
the calculations of McEachran {\it et al.} \cite{McEachran} ($-5.30 a_0$), 
and Nakanishi and Schrader \cite{Nakanishi} ($-5.09 a_0$). 
A somewhat worse agreement in N$_2$ can partially result from poorer statistics of experimental data.
Also ab-initio theoretical calculation in N$_2$ show a big spread in determination of $a_s$, see for instance 
\cite{Gianturco}.
Both for Ar and N$_2$ the model shows importance of $p$-wave scattering above 0.7-0.8 eV. 
The $s$-wave effective range for Ar amounts to $1.06 a_0$ while in
N$_2$ to $2.78 a_0$ . We recall the "size" of the N$_2$ molecule by the experimental determination
of the maximum HOMO density along the molecule axis which is about $2.3a_0$ \cite{Itatani}.
Finally from Figs.~\ref{fig:Ar} and \ref{fig:N2}
we observe that expansion into momentum series \eqref{Exp1}-\eqref{Exp3} works only at very low energies 
below $0.1$eV.
%%%%%%%%%%%%%%%%%% Table 1 %%%%%%%%%%%%%%%%%%%%%%
\begin{table}
\begin{ruledtabular}
\begin{tabular}{lllllll}
 & $R^{\ast}$($a_0$) & $E^{\ast}$(eV) & $a_\mrm{s}/R^{\ast}$ & $b$ & $R_0/R^{\ast}$ & $R_1/R^{\ast}$ \\
\hline
Ar & 3.351 & 1.211 & -1.665 & -5.138 & 0.3165 & 2.281 \\
N$_2$ & 3.397 & 1.179 & -2.729 & -12.65 & 0.8186 & --- \\
\end{tabular}
\end{ruledtabular}
\caption{\label{Tab} Characteristic distance $R^{\ast}$, characteristic energy $E^{\ast}$ and four fitting parameters:
$a_\mrm{s}$ 
($s$-wave scattering length), $R_0$ ($s$-wave effective range), $b$ (zero-energy contribution B(0) for $p$-wave)
and $R_1$ ($p$-wave effective range) for Ar and N$_2$.
}
\end{table}
%%%%%%%%%%%%%%%%%%%%%%%%%%%%%%%%%%%%%%%%%%%%%%%%%%%
%%%%%%%%%%%%%%%%%% Figure 1 %%%%%%%%%%%%%%%%%%%%%%
%%%%%%%%%%%%%%%%%%%%%%%%%%%%%%%%%%%%%%%%%%%%%%%%%%%
%%%%%%%%%%%%%%%%%% Figure 2 %%%%%%%%%%%%%%%%%%%%%%
%%%%%%%%%%%%%%%%%%%%%%%%%%%%%%%%%%%%%%%%%%%%%%%%%%%

In conclusion, we performed MERT analysis using analytical solution of Schr\"odinger equation for polarization
potential and we apply it to positron Ar (and N$_2$) scattering up to 2eV. 
The scattering length in Ar agrees well with other predictions and the effective range (for $s$-wave) is
$1.06a_0$.
More experiments are needed at low energies
to validate better the effective range parameters.

\appendix

\section{}

To solve radial Schr\"odinger equation \eqref{RadSchr} we substitute $r= \sqrt{R^\ast} e^{-z} / \sqrt{k}$ and 
$\Psi(r) = \psi(r) \sqrt{R^{\ast}/r}$, which yields the Mathieu's modified differential equation 
\cite{Erdelyi,Abramowitz}
\begin{equation}
\label{Mathieu}
\frac{d^2 \psi}{d z^2} - \left[ a - 2 q \cosh 2 z \right] \psi = 0.
\end{equation}
where $a=(l+{\tst \frac 12})^2$ and $q=k R^{\ast}$. Two 
linearly indepent solutions $M(z)$ and $T(z)$ can be expressed in the following form \cite{Erdelyi,Spector}
\begin{eqnarray}
\label{DefM}
M_\nu(z) & = & \sum_{n=-\infty}^{\infty} (-1)^n c_n(\nu) J_{2n + \nu} (2 \sqrt{q} \cosh z), \\
\label{DefT}
T_\nu(z) & = & \sum_{n=-\infty}^{\infty} (-1)^n c_n(\nu) Y_{2n + \nu} (2 \sqrt{q} \cosh z),
\end{eqnarray}
which defines them for $z>0$. Here, 
$\nu$ denotes the characteristic exponent, and $J_{\nu}(z)$ and $Y_{\nu}(z)$ are Bessel and Neumann functions
respectively. Substituting the ansatz \eqref{DefM} and \eqref{DefT} into \eqref{Mathieu} one obtains the recurrence
relation: 
\begin{equation}
\label{rec}
\left[(2n+\nu)^2 - a \right] c_n + q (c_{n-1} + c_{n+1}) = 0,
\end{equation}
which can be solved in terms of continued fractions.
To this end we introduce $h_n^{+} = c_n / c_{n-1}$ 
and $h_n^{-} = c_{-n} / c_{-n+1}$ for $n > 0$, which substituted into \eqref{rec} gives the continued fractions
$h^{+}_{n} = - q/(q h^{+}_{n+1} + d_n)$, and $h^{+}_{n} = - q/(q h^{+}_{n+1} + d_n)$ with $d_n = (2n+\nu)^2 - a$.
%\begin{equation}
%\label{rec1}
%h^{+}_{n} = - \frac{q}{q h^{+}_{n+1} + d_n}, \qquad h^{-}_{n} = - \frac{q}{q h^{-}_{n+1} + d_{-n}},
%\end{equation}
%with $d_n = (2n+\nu)^2 - a$. 
In practice to find numerical values of the coefficients $c_n$ we set $h^{+}_{m}=0$ 
and $h^{-}_{m}=0$ for some, sufficiently large $m$ and calculate $h^{+}_{n}$ and $h^{-}_{n}$ up to $n=1$.
%using \eqref{rec1}. 
Characteristic exponent has to determined from Eq.~\eqref{rec} with $n=0$.

Asymptotic behavior of $M_\nu(z)$ and $T_\nu(z)$ for large $z$ follow immediately from asymptotic expansion of Bessel functions 
\begin{align}
\label{ae1}
M_\nu(z) \stackrel{z \rightarrow \infty}{\longrightarrow} & \sqrt{\frac 2 \pi} \frac{e^{-z/2}}{q^{1/4}} s_\nu
\cos \left( {\tst e^{z} \sqrt q - \frac \pi 2 \nu  - \frac \pi 4 }\right) \\
\label{ae2}
T_\nu(z) \stackrel{z \rightarrow \infty}{\longrightarrow} & \sqrt{\frac 2 \pi} \frac{e^{-z/2}}{q^{1/4}} s_\nu
\sin \left( {\tst e^{z} \sqrt q - \frac \pi 2 \nu  - \frac \pi 4 }\right) 
\end{align}
where $s_\nu = \sum_{n=-\infty}^{\infty} c_n (\nu)$. To obtain asymptotic behavior
for large and negative $z$ one has to join solutions $M_\nu(z)$ and $T_\nu(z)$, with another pair of solution $M_\nu(-z)$ and $T_\nu(-z)$ at $z=0$ \cite{Spector}. This yields 
\begin{align}
\label{ae3}
M_\nu(z) \stackrel{z \rightarrow - \infty}{\longrightarrow} & \sqrt{\frac 2 \pi} \frac{e^{z/2}}{q^{1/4}} m s_\nu
\cos \left( {\tst \sqrt q  e^{-z} + \frac \pi 2 \nu  - \frac \pi 4 }\right) \\
T_\nu(z) \stackrel{z \rightarrow - \infty}{\longrightarrow} & - \sqrt{\frac 2 \pi} \frac{e^{z/2}}{q^{1/4}} 
\frac{s_\nu}{m} 
\Big[ \sin \left( {\tst \sqrt q e^{-z} + \frac \pi 2 \nu  - \frac \pi 4 }\right) \nonumber \\
\label{ae4}
& - \cot \pi \nu (m^2-1) \cos \left( {\tst \sqrt q e^{-z} + \frac \pi 2 \nu  - \frac \pi 4 }\right) \Big]
\end{align} 
where $m = \lim_{z \rightarrow 0^{+}} M_\nu(z) / M_{-\nu}(z)$. 

Finally we write the wave function $\Psi(r)$ in the form 
\begin{align}
\label{comb}
\Psi(r) = & \sin ({\tst \phi + \frac \pi 2 \nu  + \frac \pi 4}) \sqrt{\frac{R^{\ast}}{r}}
M_\nu\left( \ln \frac{ \sqrt{R^{\ast}}}{\sqrt{k} r} \right) \nonumber  \\
& + \cos ({\tst \phi + \frac \pi 2 \nu  + \frac \pi 4}) \sqrt{\frac{R^{\ast}}{r}}
T_\nu\left( \ln \frac{ \sqrt{R^{\ast}}}{\sqrt{k} r} \right),
\end{align}
where $\phi$ is a parameter which appear in the small $r$ expansion \eqref{Psi1}.
Now, the behavior of $\Psi(r)$ at small and large distance described by Eqs.~\eqref{Psi1}-\eqref{taneta}, can be readily
obtained from asymptotic expansions \eqref{ae1}-\eqref{ae4}.

\end{document}